\documentclass[11pt]{article}
\usepackage{graphicx}
\begin{document}
\title{\bf  The maximal axial parameters in equivalent
            parametrizations of crystal-field Hamiltonians of
            tetragonal and cubic symmetries}
\author{\bf J. Mulak$^{1}$, M. Mulak$^{2}$ and R. Gonczarek$^{2}$}
\date{{\it  $^{1}$ Trzebiatowski Institute of Low Temperature
            and Structure Research,\\
            Polish Academy of Sciences, 50--950, PO Box 1410,
            Wroclaw, Poland\\
            $^{2}$ Institute of Physics,
            Wroclaw University of Technology,\\
            Wyb. Wyspianskiego 27,
            50--370 Wroclaw, Poland}}
\maketitle
\begin{abstract}
\noindent The variation ranges of the axial $B_{k0}$ crystal-field parameters, for $k=2,4,6$, of tetragonal
including cubic crystal-field Hamiltonians ${\cal H}_{\rm CF}$ for all possible orientations of the relevant
reference frame are studied. The distinguished $z$-axis directions fixed by the maximal absolute values of
$B_{k0}$ are analyzed. The diagrams for any tetragonal ${\cal H}_{\rm CF}$ parametrization depicting the maximal
values of $\frac{|B_{k0}|}{M_{k}}$, where $M_{k}$ is the $2^{k}$-pole modulus, as a function of the
$x=\frac{B_{k4}}{B_{k0}}$ or $\frac{B_{64}}{B_{60}}$ ratios, together with the distinguished directions are
presented. The $\frac{max|B_{k0}|}{M_{k}}$ magnitudes and the relevant distinguished directions are the
discriminants of all the equivalent parametrizations. They vary within the intervals $(0.7395,1]$, $(0.6074,1]$
for tetragonal $k=4$ and tetragonal $k=6$  ${\cal H}_{\rm CF}$ components, respectively. Such specified
directions determine the mutual spatial orientation of the component $2^{k}$-poles of the ${\cal H}_{\rm CF}$,
and due to their rigid coupling in the ${\cal H}_{\rm CF}$, they also refer to the global ${\cal H}_{\rm CF}$
parametrization. This approach demonstrates the difference in fitting capability between the real and complex
isomodular ${\cal H}_{\rm CF}$ parametrizations.
\end{abstract}
\noindent
{\it PACS}: 71.70.Ch \\
\noindent {\it Key words}: Crystal--field Hamiltonian; Axial crystal-field parameters;
Tetragonal, cubic, crystal-field potential\\

\newpage

\section*{1. Introduction}
In the paper by the equivalent ${\cal H}_{\rm CF}$ parametrizations we understand those referring to the same
real crystal-field (CF) potential but expressed in variously orientated reference frames. Therefore, they can be
reduced to the identical form through the appropriate rotations of the involved reference systems. However, the
problem which generally remains, especially in the lower symmetry cases, is the lack of knowledge of the initial
reference frame orientation in which the parametrization has been defined. The equivalent parametrizations can
be identified and distinguished by their maximal axial parameters $B_{k0}$ or their maximal absolute values,
max$|B_{k0}|$, achieved in the specified reference frames. In general, max$|B_{k0}|\leq
M_{k}=\left\{\sum\limits_{q}|B_{kq}|^{2}\right\}^{1/2}$ reach the $M_{k}$ values only under certain conditions
for the crystal-field parameters (CFPs) ratios. It has been shown that the max$\frac{|B_{k0}|}{M_{k}}$
magnitudes turn out to be useful discriminants of the equivalent parametrizations [1]. In the case of CF
potentials of tetragonal and cubic symmetries expressed in their symmetry-adapted reference frames, there is no
difficulty to establish such an initial frame orientation taking the four-fold axis as the $z$-axis. The only
ambiguity resulting from the double choice of the $x$ and $y$ axes directions manifests itself solely in the
change of sign of the tetragonal $B_{k \pm 4}$ CFPs [2]. The absolute values of the
$|x_{44}|=|\frac{B_{44}}{B_{40}}|$ and $|x_{64}|=|\frac{B_{64}}{B_{60}}|$ ratios appear to be sufficient for the
complete and explicit description of the parameterizations.

There are in sum twelve point symmetry groups of the tetragonal and cubic symmetries. The parametrizations of
the corresponding ${\cal H}_{\rm CF}$s split into two kinds [3,4,5]. The first concerns the tetragonal I [6]
($D_{4}$, $C_{4v}$, $D_{2d}$, $D_{4h}$) and the cubic ($T$, $T_{h}$, $O$, $T_{d}$, $O_{h}$) point symmetries
with purely real CFPs: $5$ and $2$, respectively. The second concerns the tetragonal II [6] ($C_{4}$, $S_{4}$,
$C_{4h}$) point symmetries which need nominally $7$ CFPs, including the two pairs of complex ones.

In compliance with the tetragonal point symmetry of the central ion its ${\cal H}_{\rm CF}$ parametrizations can
include at most one pair of complex-conjugate CFPs per component multipole (except the quadrupole): $B_{44}$,
$B_{4-4}$; $B_{64}$, $B_{6-4}$. It means that the component multipoles parametrizations treated separately can
always be reduced to a real three-parameter form. This reduction can be achieved by the appropriate reference
system rotation about its $z$-axis [7].

Since the individual multipoles are characterized by the separate irreducible representations of the
three-dimensional rotation group, the maximal $B_{k0}$ variation ranges and the relevant distinguished
directions are analyzed for each multipole individually. So, we use the real form of the ${\cal H}_{\rm CF}$
parametrizations as the initial one. The more general complex parametrizations mentioned above are automatically
included in the analytical approach based on continuous rotations of the reference frame. Formulating the global
${\cal H}_{\rm CF}$ an additional problem can arise when all the three effective multipoles have to be expressed
simultaneously in a common reference frame (either known or a nominal one [8]). Then a pair of $B_{k4}$,
$B_{k-4}$ CFPs ($k=4$ or $6$) can occur in the tetragonal ${\cal H}_{\rm CF}$ parametrizations and an additional
degree of freedom (the phase of the complex $B_{k4}$ CFP) is introduced into the fitting procedure. This problem
is discussed in subsection 3.4 based on a classical example of $S_{4}$ symmetry.

Rotating the initial reference frame by the $(\alpha,\beta,0)$ Euler angles within the ranges that cover the
complete variation of the axial CFPs, two-dimensional functions $B_{k0}(\alpha,\beta)$ are obtained. These
graphs or their projections (contour maps) are characterized by the unique arrays of certain points further
called the particular points, at which both the partial derivatives $\partial B_{k0}/\partial\beta$ and
$\partial B_{k0}/\partial\alpha$ simultaneously equal zero [1]. These particular points can be classified as
maxima, minima or saddle points by means of the higher derivatives either directly from the contour-line maps or
from the three-dimensional $B_{k0}(\alpha,\beta)$  graphs. It is recommended to present the maps in the
normalized and dimensionless form dividing all their values by $M_{k}$. According to the orientation of the
parametrized CF potential with respect to the initial reference frame, the particular points arrange themselves
into the characteristic constellations. The $\frac{B_{k0}}{M_{k}}$ values at these points can be degenerate. The
maximum values, $\frac{{\rm max}|B_{k0}|}{M_{k}}$, are achieved in one or in several particular points. The
plotted functions $\frac{{\rm max}|B_{40}|}{M_{4}}$ $(x_{44})$, $\frac{{\rm max}|B_{60}|}{M_{6}}$ $(x_{64})$
allow one to find directly the maximal values and the corresponding distinguished directions for any $x$ ratios.
Additionally, these practical schemes give also the values and directions for all the remaining particular
points on the maps. It is worth emphasizing that the diagrams have a general character and make allowance for
the explicit analysis of any tetragonal ${\cal H}_{\rm CF}$ parametrization. Since the diagrams are of general
character all the relevant dimensionless values are given with a fairly high accuracy of order of $10^{-4}$.

\section*{2. General formulation}
Throughout the paper the tensor notation (by Wybourne [9]) for the ${\cal H}_{\rm CF}$ parametrizations is
consistently used. And so, in the case of the tetragonal I symmetries the parametrization has the form:
\begin{eqnarray}
{\cal H}_{\rm CF}({\rm tetra}\;{\rm I})&=& B_{20}C_{0}^{(2)}+ B_{40}C_{0}^{(4)}+ B_{44}\left(C_{4}^{(4)}+
            C_{-4}^{(4)}\right)+ B_{60}C_{0}^{(6)}+ B_{64}\left(C_{4}^{(6)}+
            C_{-4}^{(6)}\right),
\end{eqnarray}
whereas in the case of the tetragonal II symmetries the $B_{k\pm 4}$ are complex. In short,
$C_{q}^{k}=\sum\limits_{i}C_{q}^{k}(\theta_{i},\varphi_{i})$ where $C_{q}^{k}(\theta_{i},\varphi_{i})$ are the
normalized spherical harmonics [4,5,9] and the summation runs over $i$ denoting all the open-shell central-ion
electrons of the polar coordinates $(\theta_{i},\varphi_{i})$.

Here we investigate the $B_{k0}$ variation with the $z$-axis direction defined in an arbitrary reference frame
by only two Euler angles $(\alpha,\beta,0)$ with respect to the initial frame [10,11], where the third angle
$\gamma$ of the rotation about the final $z$-axis may be assumed to be zero as an arbitrary in the case. The
transformed $B_{k0}^{\prime}$ are related to the original $B_{k0}$ as [1,5,10,11]
\begin{equation}
B_{k0}^{\prime}=
\sum\limits_{q=-k}^{k}{\cal D}_{0q}^{(k)}(\alpha,\beta,0)B_{kq}=
\sum\limits_{q=-k}^{k}C_{q}^{(k)}(\beta,\alpha)B_{kq},
\end{equation}
where ${\cal D}_{0q}^{(k)}(\alpha,\beta,0)=C_{q}^{(k)}(\beta,\alpha)$ are the middle row matrix elements of the
${\cal D}^{(k)}(\alpha,\beta,0)$ rotation matrices [10,11]. Below in the paper $B_{kq}$ correspond to the
original reference frame, whereas their primed counterparts $B_{kq}^{\prime}$ stand for the rotationally
transformed ones.

Naturally, all the equivalent ${\cal H}_{\rm CF}$ parametrizations have to be characterized by the same
variation ranges of their CFPs including the axial ones, in other words by the ranges with the same upper and
lower limits. These particular maximal and minimal values can serve as useful discriminants for the whole
classes of the equivalent parametrizations. In order to standardize the equivalent parametrizations the
$\frac{{\rm max}|B_{k0}^{\prime}|}{M_{k}}$ magnitudes can be recommended as the best discriminants.
Consequently, the direction of the transformed reference frame $z$-axis relevant to these discriminants are now
the common distinguished directions for all the equivalent parametrizations. This approach is especially useful
in the case of low-symmetry CF potentials [1].

The axial CFPs play an extraordinary role since they are real and invariant relative to rotations about the
$z$-axis. They are also independent of the $z$-axis inversion, whereas the remaining CFPs become then inevitably
transformed according to the rule $B_{kq}\rightarrow B_{k-q}$, as the effect of the rotation by $(0,\pi,0)$.

The relationship between $B_{k0}^{\prime}$ CFPs and the two rotation angles $(\alpha,\beta,0)$ of the reference
frame can be expressed in the form of two-dimensional contour maps $\frac{B_{k0}^{\prime}(\alpha,\beta)}{M_{k}}$
with iso-$\frac{B_{k0}^{\prime}}{M_{k}}$ lines joining the points of identical $\frac{B_{k0}^{\prime}}{M_{k}}$
values. For the equivalent parametrizations such maps can be always transformed into each other. The whole
ranges of the rotation angles variation $0\leq \alpha \leq 2\pi$ and $0\leq \beta \leq \pi$ can be limited
according to the ${\cal H}_{\rm CF}$ symmetry. For the tetragonal symmetries (I and II) with the $z$-axis along
the four-fold axis the maps are periodical functions of $\alpha$ with the period of $\pi/2$. Furthermore, since
$B_{k0}^{\prime}$ CFPs are the same at points $(\alpha,\beta)$ and $(\alpha+\pi,\pi-\beta)$, what results from
their independence of the $z$-axis orientation, the range of $\beta$ can be confined to the interval
$(0,\pi/2)$. For the clarity reasons the maps are presented within the ranges $0\leq \alpha \leq \pi$ and $0\leq
\beta \leq \pi$.

As the paper concerns exclusively the rotational transformations of the axial $B_{k0}$ CFPs, only the middle
rows of the relevant rotation matrices ${\cal D}^{(k)}(\alpha,\beta,0)$ [10,11] are employed. All the non-axial
CFPs transform themselves according to the remaining rows [5,10,11]. After the initial reference frame rotation
the ${\cal H}_{\rm CF}$ form obeys the central-ion point symmetry with respect to the new frame. Additionally,
in the distinguished reference frames some correlations between the CFPs can occur, like that found by Burdick
and Reid [12]. They proved analytically that the $B_{k\pm1}$ CFPs become zero at the maximum, minimum and saddle
points of the $B_{k0}(\alpha,\beta)$ functions.

\section*{3. The axial components in equivalent parametrizations of tetragonal and cubic ${\cal H}_{\rm CF}$s as
functions of $\frac{B_{44}}{B_{40}}$ or  $\frac{B_{64}}{B_{60}}$ ratios}
\subsection*{3.1 The $2^{2}$-pole axial parameter $B_{20}$ in arbitrarily orientated reference frame}

In cubic potentials the quadrupolar moment is completely compensated and, in consequence, its components do not
appear in any reference frame. Hence, in the most general case the cubic ${\cal H}_{\rm CF}$ can be parametrized
by at most 22 instead of 27 CFPs.

For the tetragonal potential expressed in the symmetry-adapted system there exists only the axial term
$B_{20}C_{0}^{(2)}$. In any other reference frame rotated by $\alpha$ and $\beta$ one gets [1,10,11]:
\begin{equation}
B_{2q}^{\prime}=B_{20}{\cal D}_{q0}^{(2)}(\alpha,\beta,0)=(-1)^{q}C_{q}^{(2)}(\beta,\alpha)B_{20},
\end{equation}
and the ratios $\frac{B_{2q}^{\prime}}{B_{20}}$ for different $q$ are correlated. For $q=0$
\begin{equation}
B_{20}^{\prime}=\frac{1}{2}\left(3c^{2}-1\right)B_{20},
\end{equation}
where $c=\cos\beta$ and thus $B_{20}^{\prime}$ depends solely on $\beta$. The derivative $\partial
B_{20}^{\prime}/\partial\beta=-3csB_{20}$, where $s=\sin\beta$, and is equal to zero for $\beta=0,\pi$ or
$\pi/2$. For these $\beta$ values $B_{20}^{\prime}$ reaches its extreme magnitudes: $B_{20}$ and
$-\frac{1}{2}B_{20}$, respectively. Hence, the distinguished direction for which
$\frac{|B_{20}^{\prime}|}{M_{2}}$ achieves $1$ is the symmetry axis, whereas in the symmetry adapted system
$\frac{B_{20}}{M_{2}}$ equals 1 automatically. For $\beta=\arccos\frac{1}{\sqrt{3}}$, which is the angle between
the edge and diagonal in a cube, $B_{20}^{\prime}$ vanishes, but then the four remaining CFPs appear:
$|B_{22}^{\prime}|=|B_{2-2}^{\prime}|=\frac{\sqrt{6}}{6}B_{20}$ and
$|B_{21}^{\prime}|=|B_{2-1}^{\prime}|=\frac{\sqrt{3}}{3}B_{20}$.
\subsection*{3.2 The $2^{4}$-pole axial parameter $B_{40}$ in arbitrarily orientated reference frame}
The fourth-order ${\cal H}_{\rm CF}$ term in the tetragonal I and cubic CF potentials described in the
symmetry-adapted system has the form [3-5,13]
\begin{equation}
{\cal H}_{\rm CF}^{(IV)}=B_{40}\left[ C_{0}^{(4)}+x_{4}\left(C_{4}^{(4)}+C_{-4}^{(4)}\right) \right],
\end{equation}
and the modulus of the $2^{4}$-pole is equal to
\begin{equation}
M_{4}=|B_{40}|(1+2x_{4}^{2})^{1/2}.
\end{equation}
For the cubic case $x_{4}=\sqrt{\frac{5}{14}}=0.5976$ [6,7,13]. Eq.(3) yields for any rotation by $\alpha$ and
$\beta$ the relation:
\begin{equation}
B_{40}^{\prime}=\frac{1}{8} \left[ 3-30c^{2}+35c^{4}+x_{4}\sqrt{70}s^{4}\cos 4\alpha \right] B_{40}.
\end{equation}
Therefore,
\begin{equation}
\frac{B_{40}^{\prime}}{M_{4}}= \frac{3-30c^{2}+35c^{4}+x_{4}\sqrt{70}s^{4}\cos 4\alpha}{
8(1+2x_{4}^{2})^{1/2}}\cdot \frac{B_{40}}{|B_{40}|}.
\end{equation}
Using Eq.(9) for any $x_{4}$ the map $\frac{B_{40}^{\prime}}{M_{4}}(\alpha,\beta)$ can be depicted within the
required ranges of $\alpha$ and $\beta$ angles. The partial derivatives
\begin{eqnarray}
\frac{1}{M_{4}} \frac{\partial B_{40}^{\prime}}{\partial\beta}&=& \frac{cs(15-35c^{2}+x_{4}\sqrt{70}s^{2}\cos
4\alpha)}{ 2(1+2x_{4}^{2})^{1/2}}\cdot \frac{B_{40}}{|B_{40}|}
\nonumber\\
\frac{1}{M_{4}} \frac{\partial B_{40}^{\prime}}{\partial\alpha}&=& -\frac{x_{4}\sqrt{70}s^{4}\sin 4\alpha}{
2(1+2x_{4}^{2})^{1/2}}\cdot \frac{B_{40}}{|B_{40}|}
\end{eqnarray}
simultaneously vanish in the following five cases:
\begin{eqnarray}
&a)& s=0,
\nonumber\\
&b)& c=0, \quad \cos4\alpha=1,
\nonumber\\
&c)& c=0, \quad \cos4\alpha=-1,
\nonumber\\
&d)& 15-35c^{2}+x_{4}\sqrt{70}s^{2}=0, \quad \cos4\alpha=1,
\nonumber\\
&e)& 15-35c^{2}+x_{4}\sqrt{70}s^{2}=0, \quad \cos4\alpha=-1,
\end{eqnarray}
which correspond to the following points on the map $\frac{B_{40}^{\prime}}{M_{4}}(\alpha,\beta)$
\begin{eqnarray}
&a)& \beta=0,
\nonumber\\
&b)& \beta=\frac{\pi}{2}, \quad \alpha=2n\frac{\pi}{4},
\nonumber\\
&c)& \beta=\frac{\pi}{2}, \quad \alpha=(2n+1)\frac{\pi}{4},
\nonumber\\
&d)& \beta=\arccos \left(\frac{x_{4}\sqrt{70}+15}{x_{4}\sqrt{70}+35}\right)^{1/2}, \quad \alpha=2n\frac{\pi}{4}
\nonumber\\
&e)& \beta=\arccos \left(\frac{x_{4}\sqrt{70}-15}{x_{4}\sqrt{70}-35}\right)^{1/2}, \quad
\alpha=(2n+1)\frac{\pi}{4}.
\end{eqnarray}
In Eqs (11) $n$ stands for any integer. Let us denote by $|B_{40}^{(i)}|$, $i=a\ldots e$, the absolute value of
$B_{40}$ for the $(i)$-th condition in Eq.(10). Since the plots $\frac{|B_{40}^{(i)}|}{M_{4}}(x_{4})$ for $(a)$
and for the pairs $(b,c)$ and $(d,e)$ are symmetrical with respect to $x_{4}$ (Eqs 8, 11), only the positive
values of $x_{4}$ are taken into account.

For each condition (Eqs 10, 11) the expressions $\frac{|B_{40}^{(i)}|}{M_{4}}(x_{4})$ can be represented by the
following plots (Eq. 8):
\begin{itemize}
\item[a)]
$\frac{|B_{40}^{(a)}|}{M_{4}}(x_{4})=\left(1+2x_{4}^{2}\right)^{-1/2}$\\
For $x_{4}=0$, i.e. for the purely axial ${\cal H}_{\rm CF}$, this function reaches maximum $1$. For the cubic
case ($x_{4}=0.5976$) it amounts to $0.7638$, and monotonically falls to zero when $x_{4}\rightarrow\infty$
(Fig. 2).
\item[b)]
$\frac{|B_{40}^{(b)}|}{M_{4}}(x_{4})=
\frac{3+x_{4}\sqrt{70}}{8\left(1+2x_{4}^{2}\right)^{1/2}}$\\
For $x_{4}=0$ this function equals to $0.3750$. For the coplanar square ratio ($x_{4}=1.3944$) it reaches its
maximum $0.8292$, and then asymptotically decreases to $0.7395$ when $x_{4}\rightarrow\infty$.
\item[c)]
$\frac{|B_{40}^{(c)}|}{M_{4}}(x_{4})=
\frac{|3-x_{4}\sqrt{70}|}{8\left(1+2x_{4}^{2}\right)^{1/2}}$\\
This function for $x_{4}=0$ takes the same value as in (b) since they are symmetrical. When $x_{4}$ rises it
drops to zero at $x_{4}=0.3586$, and then monotonically increases to $0.7395$ for $x_{4}\rightarrow\infty$.
\item[d)]
The condition $0\leq c^{2} \leq 1$, where $c^{2}=\frac{x_{4}\sqrt{70}+15}{x_{4}\sqrt{70}+35}$, is fulfilled for
all $x_{4}\geq 0$ because $x_{4}\geq -1.7928$ is sufficient. Substituting the above expression for $c^{2}$ and
$\cos4\alpha=1$ into Eq.8 yields\\
$\frac{|B_{40}^{(d)}|}{M_{4}}(x_{4})= \frac{|x_{4}\sqrt{70}-15|}{
(x_{4}\sqrt{70}+35)\left(1+2x_{4}^{2}\right)^{1/2}}$\\
For $x_{4}=0$ the function equals $0.4286$, it has zero-point at $x_{4}=1.7928$ and then it pursues zero for
$x_{4}\rightarrow\infty$ passing through the weak and flat maximum $0.0489$ at $x_{4}=5.1532$.
\item[e)]
The condition $0\leq c^{2} \leq 1$, where $c^{2}=\frac{x_{4}\sqrt{70}-15}{x_{4}\sqrt{70}-35}$, confines its
domain of determinancy to \linebreak $x_{4}\leq1.7928$.
Within this range\\
$\frac{|B_{40}^{(e)}|}{M_{4}}(x_{4})=\frac{x_{4}\sqrt{70}+15}{
(|x_{4}\sqrt{70}-35|)\left(1+2x_{4}^{2}\right)^{1/2}}$.\\
For $x_{4}=0$ this function equals $0.4286$, and for the limiting value $x_{4}=1.7928$ it is $0.5504$, i.e. the
same as in $(c)$ at that point. Within the interval $0\leq x_{4}\leq 1.7928$ the plot passes through the subtle
local maximum of $0.5092$ at $x_{4}=0.5977$, and minimum $0.5065$ at $x_{4}=0.9701$.
\end{itemize}
The plots $\frac{|B_{40}^{(i)}|}{M_{4}}(x_{4})$, $i=a,b,c,d$ and $e$ corresponding to the five conditions (Eqs
10, 11) are presented in Fig. 2. The diagram shows the positions and values of all particular points on the
$\frac{B_{40}^{\prime}}{M_{4}}(\alpha,\beta)$ map for any tetragonal ${\cal H}_{\rm CF}$ as a function of
$x_{4}$ (see Figs 5-7). Any vertical straight line drawn for a given $x_{4}$ intersects the diagram branches
(from $a$ to $e$) in three to five points (depending on the $x_{4}$) directly determining the distinguished
directions and values including the $\frac{{\rm max}|B_{40}^{\prime}|}{M_{4}}$.

As is seen from the diagram the four-fold axis, i.e. the initial $z$-axis direction ($[001]$), is the
distinguished one for $|x_{4}|\leq 0.5976$. For all the remaining $x_{4}$, i.e. $|x_{4}|\geq 0.5976$, this
distinguished direction lies along the square diagonal $[110]$. The crossover point corresponds exactly to the
cubic $x_{4}$ value. The $\frac{{\rm max}|B_{40}^{\prime}|}{M_{4}}$ value varies within the range from $1.0000$
(for $x_{4}=0$) down to $0.7395$ for $x_{4}\rightarrow\infty$, achieving on the way a local minimum $0.7638$ for
the cubic $x_{4}$ value, and a local maximum $0.8292$ for the coplanar square $x_{4}$ value.

\subsection*{3.3 The $2^{6}$-pole axial parameter $B_{60}$ in arbitrarily orientated reference frame.}
In turn, the sixth-order  ${\cal H}_{\rm CF}$ term in the tetragonal I and cubic CF potentials in the symmetry
adapted system has the form [3-5,13]:
\begin{equation}
{\cal H}_{\rm CF}^{(VI)}=B_{60}\left[C_{0}^{(6)}+x_{6}\left(C_{4}^{(6)}+C_{-4}^{(6)}\right) \right],
\end{equation}
and the modulus of the $2^{6}$-pole is equal to
\begin{equation}
M_{6}=|B_{60}|(1+2x_{6}^{2})^{1/2}.
\end{equation}
For the cubic case $x_{6}=\sqrt{\frac{7}{2}}=1.8708$ [6,7,13]. In the reference frame rotated by
$(\alpha,\beta,0)$ the axial CFP $B_{60}$ transforms to (see Eq.3):
\begin{equation}
B_{60}^{\prime}=\frac{1}{16}\left[ -5+105c^{2}-315c^{4}+231c^{6}+x_{6}3\sqrt{14}s^{4}(-1+11c^{2})\cos4\alpha
\right]B_{60}.
\end{equation}
Hence,
\begin{equation}
\frac{B_{60}^{\prime}}{M_{6}}=\frac{-5+105c^{2}-315c^{4}+231c^{6}+x_{6}3\sqrt{14}s^{4}(-1+11c^{2})
\cos4\alpha}{16(1+2x_{6}^{2})^{1/2}}\cdot \frac{B_{60}}{|B_{60}|}.
\end{equation}
The particular points of the map $\frac{B_{60}^{\prime}}{M_{6}}(\alpha,\beta)$, especially the point (or points)
corresponding to the $\frac{{\rm max}|B_{60}^{\prime}|}{M_{6}}$, are those for which both the partial
derivatives $\partial B_{60}^{\prime}/\partial\beta$ and $\partial B_{60}^{\prime}/\partial\alpha$
simultaneously become zero
\begin{eqnarray}
\frac{1}{M_{6}} \frac{\partial B_{60}^{\prime}}{\partial\beta}&=&
\frac{3}{8}sc\frac{-35+210c^{2}-231c^{4}+x_{6}\sqrt{14} (-13+46c^{2}-33c^{4})\cos4\alpha}{
(1+2x_{6}^{2})^{1/2}}\cdot \frac{B_{60}}{|B_{60}|}
\nonumber\\
\frac{1}{M_{6}} \frac{\partial B_{60}^{\prime}}{\partial\alpha}&=&
-\frac{3}{4}\frac{x_{6}\sqrt{14}s^{4}(-1+11c^{2})\sin 4\alpha}{ (1+2x_{6}^{2})^{1/2}}\cdot
\frac{B_{60}}{|B_{60}|}.
\end{eqnarray}
There are six possibilities of simultaneous vanishing of the partial derivatives
\begin{eqnarray}
&a)& s=0,
\nonumber\\
&b)& c=0, \quad \cos4\alpha=1,
\nonumber\\
&c)& c=0, \quad \cos4\alpha=-1,
\nonumber\\
&d)& c^{2}=\frac{1}{11}, \quad \cos4\alpha=-\frac{0.5238}{x_{6}},
\nonumber\\
&e)& (231+33\sqrt{14}x_{6})c^{4}-(210+46\sqrt{14}x_{6})c^{2}+
     (35+13\sqrt{14}x_{6})=0, \quad \cos4\alpha=1,
\nonumber\\
&f)& (231-33\sqrt{14}x_{6})c^{4}-(210-46\sqrt{14}x_{6})c^{2}+
     (35-13\sqrt{14}x_{6})=0, \quad \cos4\alpha=-1.
\end{eqnarray}
The last two conditions, as the functions of $x_{6}$ yield the double solutions, i.e. four solutions for
$x_{6}>0.7195$ and three ones for $|x_{6}|\leq 0.7195$, so in sum there are eight or seven particular points on
the map apart from their symmetrical images due to the presence of the four-fold axis as well the invariance of
the axial CFPs with respect to the $z$-axis sense.

The conditions specified above in Eq.17 (from $a$ to $f$) correspond to the following particular points on the
map $\frac{B_{60}^{\prime}}{M_{6}}(\alpha,\beta)$:
\begin{eqnarray}
&a)& \beta=0,
\nonumber\\
&b)& \beta=\frac{\pi}{2}, \quad \alpha=2n\frac{\pi}{4},
\nonumber\\
&c)& \beta=\frac{\pi}{2}, \quad \alpha=(2n+1)\frac{\pi}{4},
\nonumber\\
&d)& \beta=72.45^{o}, \quad \alpha=\frac{1}{4} \arccos\left(\frac{-0.5238}{x_{6}}\right),
\nonumber\\
&e)& \beta=\arccos \left[\frac{1}{33}\frac{15\sqrt{14}+46x_{6} -2(100x_{6}^{2}+48\sqrt{14}x_{6}+210)^{1/2}
}{\sqrt{14}+2x_{6}}\right]^{1/2}, \quad \alpha=2n\frac{\pi}{4},
\nonumber\\
&e^{\prime})& \beta=\arccos \left[\frac{1}{33}\frac{15\sqrt{14}+46x_{6}
+2(100x_{6}^{2}+48\sqrt{14}x_{6}+210)^{1/2} }{\sqrt{14}+2x_{6}}\right]^{1/2}, \quad \alpha=2n\frac{\pi}{4},
\nonumber\\
&f)& \beta=\arccos \left[\frac{1}{33}\frac{15\sqrt{14}-46x_{6} -2(100x_{6}^{2}-48\sqrt{14}x_{6}+210)^{1/2}
}{\sqrt{14}-2x_{6}}\right]^{1/2}, \quad \alpha=(2n+1)\frac{\pi}{4},
\nonumber\\
&f^{\prime})& \beta=\arccos \left[\frac{1}{33}\frac{15\sqrt{14}-46x_{6}
+2(100x_{6}^{2}-48\sqrt{14}x_{6}+210)^{1/2} }{\sqrt{14}-2x_{6}}\right]^{1/2}, \quad \alpha=(2n+1)\frac{\pi}{4}.
\end{eqnarray}
In the cases $(d)$, $(e)$ and $(f)$ the definite domains of determinancy of the involved expressions have to be
taken into account. In consequence, the $x_{6}$ ratio has to fulfil the following conditions: $|x_{6}|\geq
0.5238$, $x_{6}\geq -0.7195$ and $x_{6}\leq 0.7195$, for $(d)$, $(e)$ and $(f)$, respectively.

Unlike for the particular points in $(a-c)$ cases their location in $(d-f)$ cases does depend on the $x_{6}$
ratio. The $(d)$ case turns to be quite intriguing. The corresponding particular points (Eq.18d) lie along the
straight line $\beta=72.45^{o}$, and the $\frac{|B_{60}^{(d)}|}{M_{6}}(x_{6})$ function takes along it the
constant value $\frac{0.1322}{(1+2x_{6}^{2})^{2}}$ regardless of the $\alpha$ rotation angle.

In the $(e)$ case the $\beta$ angle varies from $90^{o}$ for the limiting value $x_{6}=-0.7195$, through
$62.04^{0}$ for $x_{6}=0$, and monotonically goes down to $51.12^{o}$ for $x_{6}\rightarrow\infty$. The
$\beta(x_{6})$ plot for the $(f)$ case is symmetrical to the previous one with respect to $x_{6}$. Similarly,
the symmetrical plots for the $(e^{\prime})$, Eq.(18$e^{\prime})$ and $(f^{\prime})$, Eq.(18$f^{\prime})$, cases
are characterized by $\beta=51.12^{o}$ for $x_{6}\rightarrow\mp\infty$, $\beta=33.88^{o}$ for $x_{6}=0$, and
$\beta=0$ for $x_{6}\rightarrow\pm\infty$, respectively (Fig.3). Since the plots
$\frac{|B_{60}^{(i)}|}{M_{6}}(x_{6})$ for $(a)$, $(d)$ and the pairs $(b,c)$, $(e,f)$ and $(e^{\prime},
f^{\prime})$ are symmetrical with respect to $x_{6}$ (Eqs 15, 18) only the positive values of $x_{6}$ ratio are
considered.

For the conditions defined above (Eqs 17,18) the expressions $\frac{|B_{60}^{\prime}|}{M_{6}}(x_{6})$ reduce
themselves (Eq.15) to the following plots:
\begin{itemize}
\item[a)]
$\frac{|B_{60}^{(a)}|}{M_{6}}(x_{6})=(1+2x_{6}^{2})^{-1/2}$\\
For $x_{6}=0$ the maximal possible value $1$ is reached, for the cubic ratio ($x_{6}=1.8708$) it amounts to
$0.3536$, and then monotonically falls down to zero as $x_{6}\rightarrow\mp\infty$.
\item[b)]
$\frac{|B_{60}^{(b)}|}{M_{6}}(x_{6})=
\frac{|-5-3\sqrt{14}x_{6}|}{16(1+2x_{6}^{2})^{1/2}}$\\
For $x_{6}=0$ the function is equal to $0.3125$, for the coplanar square ratio ($x_{6}=1.1225$) it reaches its
maximum $0.5863$, and then slowly falls to the limit $0.4961$ as $x_{6}\rightarrow\infty$.
\item[c)]
$\frac{|B_{60}^{(c)}|}{M_{6}}(x_{6})=
\frac{|-5+3\sqrt{14}x_{6}|}{16(1+2x_{6}^{2})^{1/2}}$\\
As above, for $x_{6}=0$ the function equals $0.3125$, and when $x_{6}$ rises it falls to zero for
$x_{6}=0.4454$, then monotonically increases to the asymptotic value $0.4961$ if $x_{6}\rightarrow\infty$.
\item[d)]
$\frac{|B_{60}^{(d)}|}{M_{6}}(x_{6})=
\frac{0.1322}{(1+2x_{6}^{2})^{1/2}}$\\
Here, the domain is limited to $|x_{6}|\geq 0.5238$, (Eq.18). For $x_{6}=0.5238$ the function amounts to
$0.1062$ and monotonically tends to zero for $x_{6}\rightarrow\infty$.
\item[e)]
$\frac{|B_{60}^{(e)}|}{M_{6}}(x_{6})= \frac{|-5+105t_{1}(x_{6})-315t_{1}^{2}(x_{6})+231t_{1}^{3}(x_{6})+
x_{6}3\sqrt{14}(1-t_{1}(x_{6}))^{2}(-1+11t_{1}(x_{6}))|}{
16(1+2x_{6}^{2})^{1/2}}$,\\
where $t_{1}(x_{6})=\frac{1}{33} \frac{15\sqrt{14}+46x_{6}-2(100x_{6}^{2}+48\sqrt{14}x_{6}+210)^{1/2}}{
\sqrt{14}+2x_{6}}\;\;$, and $\;\;x_{6}\geq -0.7195$.\\
For $x_{6}=0$ the function equals $0.3321$. For the cubic ratio ($x_{6}=1.8708$) it reaches quite a subtle
maximum $0.6285$, and then slowly falls to $0.6074$ if $x_{6}\rightarrow\infty$.
\item[e$^{\prime}$)]
$\frac{|B_{60}^{(e^{\prime})}|}{M_{6}}(x_{6})=
\frac{|-5+105t_{2}(x_{6})-315t_{2}^{2}(x_{6})+231t_{2}^{3}(x_{6})+
x_{6}3\sqrt{14}(1-t_{2}(x_{6}))^{2}(-1+11t_{2}(x_{6}))|}{
16(1+2x_{6}^{2})^{1/2}}$,\\
where $t_{2}(x_{6})=\frac{1}{33} \frac{15\sqrt{14}+46x_{6}+2(100x_{6}^{2}+48\sqrt{14}x_{6}+210)^{1/2}}{
\sqrt{14}+2x_{6}}$.\\
For $x_{6}=0$ the expression equals $0.4148$, it has zero-point at $x_{6}=1.2481$, next slightly rises to the
maximum $0.0710$ at $x_{6}=3.6498$, and then slowly falls to zero if $x_{6}\rightarrow\infty$.
\item[f)]
$\frac{|B_{60}^{(f)}|}{M_{6}}(x_{6})= \frac{|-5+105t_{1}^{\prime}(x_{6})-315t_{1}^{\prime 2}(x_{6})+
231t_{1}^{\prime 3}(x_{6})- x_{6}3\sqrt{14}(1-t_{1}^{\prime}(x_{6}))^{2}(-1+11t_{1}^{\prime}(x_{6}))|}{
16(1+2x_{6}^{2})^{1/2}}$,\\
where $t_{1}^{\prime}(x_{6})=\frac{1}{33} \frac{15\sqrt{14}-46x_{6}-2(100x_{6}^{2}-48\sqrt{14}x_{6}+210)^{1/2}}{
\sqrt{14}-2x_{6}}$ and $x_{6}\leq 0.7195$.\\
For $x_{6}=0$ the function equals $0.3321$. At the limiting point ($x_{6}=0.7195$) it amounts to $0.1348$, and
within the interval $(0\leq x_{6}\leq 0.7195)$ the plot passes through the minimum $0.1050$ at $x_{6}=0.5595$.
\item[f$^{\prime}$)]
$\frac{|B_{60}^{(f^{\prime})}|}{M_{6}}(x_{6})= \frac{|-5+105t_{2}^{\prime}(x_{6})-315t_{2}^{\prime 2}(x_{6})+
231t_{2}^{\prime 3}(x_{6})- x_{6}3\sqrt{14}(1-t_{2}^{\prime}(x_{6}))^{2}(-1+11t_{2}^{\prime}(x_{6}))|}{
16(1+2x_{6}^{2})^{1/2}}$,\\
where $t_{2}^{\prime}(x_{6})=\frac{1}{33} \frac{15\sqrt{14}-46x_{6}+2(100x_{6}^{2}-48\sqrt{14}x_{6}+210)^{1/2}}{
\sqrt{14}-2x_{6}}$.\\
For $x_{6}=0$ the function equals $0.4148$ and then monotonically rises to $0.6074$ as $x_{6}\rightarrow\infty$.
\end{itemize}

\noindent The $\frac{|B_{60}^{(i)}|}{M_{6}}(x_{6})$ plots, where $i=a,b,\ldots,f^{\prime}$, for all the eight
conditions (Eqs 17,18) are presented in Fig.4. The straight line corresponding to any chosen $x_{6}$ ratio
intersects the diagram branches in four to eight points depending on the $x_{6}$ value (see Figs 5-7). The
points of intersection determine the positions of the particular points and the respective values including the
$\frac{{\rm max}|B_{60}^{\prime}|}{M_{6}}(x_{6})$ discriminant. The characteristic degeneracy of some particular
point values in the diagrams manifesting itself as the intersection points of their different branches (Figs
2,4) reflects some definite symmetry encoded in the $x_{6}$ ratio.

As is seen from the diagram (Fig.4) the four-fold axis, i.e. the initial $z$-axis direction ($[001]$), is the
distinguished direction for $|x_{6}|\leq 0.9179$. For all the remaining $x_{6}$, i.e. $|x_{6}|\geq 0.9171$, the
distinguished directions (roughly $[111]$) and the corresponding maximal values are determined by the $e$-th
branch of the diagram. In the cross-over point ($x_{6}=0.9179$) this maximal value amounts to $0.6103$. For
$x_{6}>0.9179$ the distinguished direction forms with the reference frame $z$-axis the angle $\beta$ given by
Eq.18e. In the crossover point this angle amounts to $56.69^{o}$, for the cubic ratio ($x_{6}=1.8708$) it equals
$54.74^{o}$ (exactly $[111]$ direction), and then asymptotically pursues the limit $51.12^{o}$ as
$x_{6}\rightarrow\infty$.

The $\frac{{\rm max}|B_{60}^{\prime}|}{M_{6}}(x_{6})$ can vary within the range from $1.0000$ (for $x_{6}=0$) to
$0.6074$ (for $x_{6}\rightarrow\infty$) passing on the way through the local minimum $0.6103$ (for
$x_{6}=0.9179$) and the maximum $0.6285$ (for $x_{6}=1.8708$). It is worth noticing that the distinguished
direction crossover takes place for $x_{6}=0.9179$ which is smaller than both the coplanar square ratio $1.1225$
and the cubic ratio $1.8708$.

\subsection*{3.4 Tetragonal II ${\cal H}_{\rm CF}$ with complex parameters -- the maximal axial CFPs
in arbitrarily orientated reference frame}
The tetragonal II ${\cal H}_{\rm CF}$s containing two pairs of $B_{k4}$, $B_{k-4}$ CFPs $(k=4,6)$  in the
symmetry adapted system deserves more careful analysis. If only one pair of the complex CFPs occurs in the
${\cal H}_{\rm CF}$, there always exists the equivalent real parametrization. The complete parametrization of
the tetragonal II ${\cal H}_{\rm CF}$ contains seven instead of five CFPs. Taking into account the rotational
properties of ${\cal H}_{\rm CF}$ it is recommended to use the pairs of complex-conjugate parameters:
$B_{k4}=|B_{k4}|{\rm e}^{{\rm i}4\varphi_{k4}}$ and $B_{k-4}=|B_{k4}|{\rm e}^{-{\rm i}4\varphi_{k4}}$ rather
than the equivalent pairs Re$B_{k4}$ and Im$B_{k4}$. Between the former and latter pairs of the CFPs the
following relations hold: $B_{k4}={\rm Re}B_{k4}+{\rm iIm}B_{k4},\;$ $B_{k-4}={\rm Re}B_{k4}-{\rm iIm}B_{k4},\;$
$|B_{k4}|=\left[({\rm Re}B_{k4})^{2}+({\rm Im}B_{k4})^{2}\right]^{1/2}$,
$\varphi_{k4}=\frac{1}{4}\arctan\frac{{\rm Im}B_{k4}}{{\rm Re}B_{k4}}\;$ [4,7]. Rotation about the $z$-axis (the
four-fold axis) by the angle $\varphi_{k4}$ eliminates the ${\rm Im}B_{k4}$. Then, the new real
$B_{k4}^{\prime}$ is equal to $|B_{k4}|$, and the tetragonal parameters for the remaining $k$ transform
correspondingly, whereas the axial CFPs ($q=0$) remain unchanged. Therefore, the tetragonal II ${\cal H}_{\rm
CF}$ for $d$-electron transition ions can always be parametrized purely by the real CFPs. Similarly, for the
hexagonal $C_{3h}$ symmetry the number of CFPs can be reduced from five down to four and is equal to that for
$D_{3h}$ symmetry, when the ${\rm Im}B_{66}$ is absent due to the higher symmetry [7,8,14]. Such elimination of
one of the two ${\rm Im}B_{k4}$ CFPs ($k=4$ or $6$) fixes the reference frame, and is equivalent to the lost of
one degree of freedom, i.e. the rotation of the reference frame [6]. This fixed frame is imposed on the
remaining effective multipoles in the ${\cal H}_{\rm CF}$ to be parametrized.

Let us consider as an example the ${\cal H}_{\rm CF}$ of $S_{4}$ point symmetry occurring in paramagnetic ions
in the scheelite type crystal structure [7,15]. Then, the complete seven-parameter ${\cal H}_{\rm CF}$ can be
written in the form
\begin{eqnarray}
{\cal H}_{\rm CF}(S_{4})=B_{20}C_{0}^{(2)}&+&B_{40}C_{0}^{(4)}+|B_{44}|{\rm e}^{i4\varphi_{44}}C_{4}^{(4)}+
|B_{44}|{\rm e}^{-i4\varphi_{44}}C_{-4}^{(4)}+\nonumber \\&+&B_{60}C_{0}^{(6)}+|B_{64}|{\rm
e}^{i4\varphi_{64}}C_{4}^{(6)}+ |B_{64}|{\rm e}^{-i4\varphi_{64}}C_{-4}^{(6)},
\end{eqnarray}
where the $B_{k4}$ and $B_{k-4}$ CFPs are only conjugate (Eq.(1)).

This form is usually reduced to the six-parameter one. The $B_{4-4}$ is eliminated rotating the reference frame
by the angle $\varphi_{44}$, and so we get
\begin{eqnarray}
{\cal H}_{\rm CF}(S_{4})=B_{20}C_{0}^{(2)}&+&B_{40}C_{0}^{(4)}+|B_{44}|\left(C_{4}^{(4)}+
C_{-4}^{(4)}\right)+B_{60}C_{0}^{(6)}+\nonumber \\&+&|B_{64}|{\rm e}^{i4(\varphi_{64}-\varphi_{44})}C_{4}^{(6)}+
|B_{64}|{\rm e}^{-i4(\varphi_{64}-\varphi_{44}}C_{-4}^{(6)}.
\end{eqnarray}
This formula differs from the real one of the same $M$ in one additional parameter (Eq.(1)) -- the phase
difference $(\varphi_{64}-\varphi_{44})$ of the complex CFPs, which is an extra parameter in fitting procedures.
Now the $max|B_{60}^{\prime}|$ magnitude for all possible orientations of the reference frame is the maximum of
the expression (Eq.(14))
\begin{eqnarray}
|B_{60}^{\prime}|=\frac{1}{16}\left|-5+105c^{2}-315c^{4}+231c^{6}+x_{64}3\sqrt{14}s^{4}(-1+11c^{2}) \cos 4\left[
\alpha-(\varphi_{64}-\varphi_{44})\right]\right|\cdot|B_{60}|,
\end{eqnarray}
where $x_{64}=\frac{|B_{64}|}{B_{60}}$. Thus the shift $(\varphi_{64}-\varphi_{44})$ does not change the
$max|B_{60}^{\prime}|$, but influences the relevant distinguished direction compared to that for the real
parametrization and, in consequence, modifies the spatial orientation of the considered multipole.

Let us note that in the case of the tetragonal I ${\cal H}_{\rm CF}$, i.e. for real CFPs, there is no such
possibility -- the orientation of the considered $2^{k}$-pole is absolutely determined by the $x_{k4}$-ratio.
The result is that the complex ${\cal H}_{\rm CF}$ parametrizations enable some extra modifications of the
spatial orientation of the component multipoles in fitting procedures when the absolute values of the CFPs are
conserved.

\section*{5. Discussion}
The maximal admissible magnitudes of the $B_{k0}$ CFPs, their variation ranges as well as the relevant
distinguished directions depending on the reference frame orientation for all classes of the equivalent
parametrizations for the tetragonal including cubic ${\cal H}_{\rm CF}$s have been thoroughly analyzed. The
analysis was inspired by observation that in any reference frame the entire magnitudes of $M_{k}$ cannot be
generally achieved by the parameters of the ${\cal H}_{\rm CF}$ $2^{k}$-pole component, particularly by the
axial parameters $B_{k0}$ for even $k\geq 2$. The only exception is the purely axial ${\cal H}_{\rm CF}$s for
which $\frac{max|B_{k0}^{\prime}|}{M_{k}}=1$ if the reference system $z$-axis is in line with the symmetry axis.
The quadrupole ($k=2$) is a special case (not to mention vectors with $k=1$) since $|B_{20}^{\prime}|$ can reach
$M_{2}$ for $x_{2}=B_{22}/B_{20}=\frac{\sqrt{6}}{2}$ [16], but this case does not concern the considered
symmetries.

The calculated dependencies  of the $\frac{max|B_{k0}^{\prime}|}{M_{k}}(x_{k})$ for $k=4,6$ on the ${\cal
H}_{\rm CF}$ $2^{k}$-pole composition represented by the appropriate $x$ ratio are shown in Figs 2 and 4. They
offer the general review of all particular points of the maps $\frac{B_{k0}^{\prime}}{M_{k}}(\alpha,\beta)$
establishing their positions and giving the relevant values for any $x_{4}$ and $x_{6}$ ratios. The
$\frac{max|B_{k0}^{\prime}|}{M_{k}}$ values can serve as discriminants of all the equivalent parametrizations
which have the same form in the common symmetry adapted system.

The degree of reduction of the $\frac{max|B_{k0}^{\prime}|}{M_{k}}$ discriminants with respect to $1$ is more
tangible for $k=6$ than for $k=4$. The whole ranges of the discriminants as functions of $x$ are limited. The
lower limits of $\frac{max|B_{40}^{\prime}|}{M_{4}}$ and $\frac{max|B_{60}^{\prime}|}{M_{6}}$ amount to $0.7395$
and $0.6074$, respectively, whereas for comparison in the case of quadrupole
$\frac{|B_{20}^{\prime}|}{M_{2}}\geq 0.8660$ [16]. The envelopes $\frac{max|B_{k0}^{\prime}|}{M_{k}}$ vs. $x$
are characterized by a sharp local minimum at the intersection point of the two plots representing the two
various conditions for the extremum, and a weak flat maximum for the ratios $x_{44}=1.3944$ and $x_{64}=1.8708$
corresponding to the co-planar square and cubic coordinations, respectively. These particular $x$ ratios can be
linked up with the coordination geometry. Taking into account the point symmetry of the central ion surroundings
and $1/R$ dependence of the individual ligand potential one gets [4]
\begin{eqnarray}
\frac{B_{44}}{B_{40}}&=&\pm \frac{\sqrt{70}}{2(2\cot^{4}\theta-12\cot^{2}\theta+3)},\nonumber\\
\frac{B_{64}}{B_{60}}&=&\pm
\frac{3\sqrt{14}}{2}\frac{5\cot^{2}\theta-1}{2\cot^{6}\theta-30\cot^{4}\theta+45\cot^{2}\theta-5)},
\end{eqnarray}
where $\theta$ and $R$ stand for the spherical coordinates of the ligands forming a square. The sign of the
ratios depends on the $x$ and $y$ axes orientation with respect to the square.

The distinguished directions corresponding to the maximal magnitudes of $\frac{|B_{k0}^{\prime}|}{M_{k}}(x)$ are
presented in Fig.8. All the directions are defined with respect to the reference frame in which the initial
parametrization is expressed [1,8].

The above investigations provide the distinguished directions for all the component multipoles separately, but
they are also obligatory for the resultant ${\cal H}_{\rm CF}$. These are the directions that could be used in a
possible reduction of the number of independent CFPs [12], as well as examining the standardization of
equivalent parametrizations [17,18]. The recognition of the distinguished directions for individual
$2^{k}$-poles gives their autonomous spatial orientation and throws light on the mechanism how the component
multipoles combine into the resultant effect. The role of the mutual orientation of the component multipoles is
clearly noticeable while the real (five-parameter) tetragonal parametrizations of ${\cal H}_{\rm CF}$ are
compared with the more general complex (seven- or six-parameter) iso-modular parametrizations. The latter is
more capable in fitting procedures what implies directly from the possibility of modification of the component
multipoles mutual orientation. Dealing with the lower symmetry ${\cal H}_{\rm CF}$, in which the component
multipoles contain more than one complex-conjugate pairs of CFPs, some intra-multipoles modifications are also
possible.

The $\frac{B_{40}^{\prime}}{M_{4}}(\alpha,\beta)$ and $\frac{B_{60}^{\prime}}{M_{6}}(\alpha,\beta)$ maps for
three representative, well-documented parametrizations of the tetragonal ${\cal H}_{\rm CF}$s significantly
differing in $x_{44}$ and $x_{64}$ ratios are shown in Figs 5-7 as an example. The maps in Fig.5 show the
variation of the $B_{40}$ and $B_{60}$ CFPs depending on the reference frame $z$-axis direction for the
Pr$^{3+}$ ion embedded in the KY$_{3}$F$_{10}$ matrix [19]. Its ${\cal H}_{\rm CF}$ of $C_{4v}$ point symmetry
is characterized by the following parameters (in $cm^{-1}$): $B_{20}=-589$, $B_{40}=-1711$, $B_{44}=519$,
$B_{60}=657$, $B_{64}=-196$; $M_{2}=589$, $M_{4}=1862$, $M_{6}=713$; $x_{4}=-0.303$, $x_{6}=-0.298$.

The maps in Fig.6, in turn, apply to the cubic ${\cal H}_{\rm CF}$ with the characteristic ratios
$x_{4}=0.5976$, and $x_{6}=-1.8708$ [3-5,13]. The system Pr$^{3+}$:Cs$_{2}$NaRCl$_{6}$ is given here as an
example, and, the $O_{h}$ CF potential is defined by the CFPs (in $cm^{-1}$): $B_{40}=2326$, $B_{60}=247$;
$M_{4}=3045$, $M_{6}=699$ [20].

Successively, the maps in Fig.7 refer to the $B_{40}$ and $B_{60}$ CFPs variation for Pr$^{+3}$ ion in
Pr$_{2}$CuO$_{4}$ [21], i.e. in the $C_{4v}$ CF potential of the ratios $x_{4}=-0.830$, $x_{6}=9.99$, and the
CFPs (in $cm^{-1}$) are correspondingly equal to: $B_{20}=-294$, $B_{40}=-2302$, $B_{44}=1910$, $B_{60}=147$,
$B_{64}=1469$; $M_{2}=294$, $M_{4}=2991$, $M_{6}=1476$. With the exception of the cubic case the remaining
$x_{4}$ and $x_{6}$ ratios come from fittings experimental data.

The particular points marked on the maps (Figs 5--7) by the letters are also noticeable on the diagrams (Figs 2
and 4) as the intersection points of the appropriate diagram branches with the vertical straight lines
corresponding to the involved $x_{4}$ and $x_{6}$ ratios. The intersection points with the highest branches
refer to the $\frac{{\rm max}|B_{k0}^{\prime}|}{M_{k}}$ discriminants.

Figs 2 and 4 present the variation of the particular points location on the maps
$\frac{B_{k0}^{\prime}}{M_{k}}(\alpha,\beta)$ with $x_{4}$ and $x_{6}$. Since the
$\frac{B_{k0}^{\prime}}{M_{k}}(\alpha,\beta)$ maps can be considered as the representations of the parent
parametrizations, further conclusions concerning the links and genealogy of various CF potentials, their optimal
higher-symmetry approximations, etc. may be drawn based on the evolution of such maps while descending in the
central ion symmetry, what can be simulated by its shifts inside the coordination polyhedron in various
directions. Although the results presented in this paper concern the equivalent ${\cal H}_{\rm CF}$
parametrizations, from the point of view of the maximal axial parameters of component multipoles and their
relevant distinguished directions they have more general character going beyond the crystal-field approach.



\newpage

\noindent
{\large\bf FIGURE CAPTIONS}

\noindent {\bf FIG.1} The dependence of $\frac{B_{20}^{\prime}}{|B_{20}|}$ on the reference frame rotation angle
$\beta$ for tetragonal ${\cal H}_{\rm CF}$s (Eq.4).

\noindent {\bf FIG.2} $\frac{|B_{40}^{(i)}|}{M_{4}}$ vs $|x_{4}|$ plots, where $i=a,b,c,d,e$ denotes one out of
the five conditions (Eq.10) or the particular points of the map $\frac{B_{40}^{\prime}(\alpha,\beta)}{M_{4}}$,
(Eq.11). The vertical straight lines refer to: $|x_{4}|=0.303$ for Pr$^{3+}$:KY$_{3}$F$_{10}$ [19];
$|x_{4}|=0.598$ for Pr$^{3+}$:Cs$_{2}$NaRCl$_{6}$ [20], and $|x_{4}|=0.830$ for Pr$_{2}$CuO$_{4}$ [21],
respectively.

\noindent {\bf FIG.3} $\beta(x_{6})$ functions for $(e)$ and $(e^{\prime})$ conditions (Eq.18); for $(f)$ and
$(f^{\prime})$ conditions the plots are symmetrical with respect to $x_{6}$.

\noindent {\bf FIG.4} $\frac{|B_{60}^{(i)}|}{M_{6}}$ vs $|x_{6}|$ plots, where
$i=a,b,c,d,e,e^{\prime},f,f^{\prime}$ denotes one out of the eight conditions (Eq.17) or the particular points
of the map $\frac{B_{60}^{\prime}(\alpha,\beta)}{M_{6}}$, (Eq.18). The vertical straight lines refer to:
$|x_{6}|=0.298$ for Pr$^{3+}$:KY$_{3}$F$_{10}$ [19] and $|x_{6}|=1.871$ for Pr$^{3+}$:Cs$_{2}$NaRCl$_{6}$ [20];
for Pr$_{2}$CuO$_{4}$ $x_{6}=9.99$ [21].

\noindent {\bf FIGS 5a, 5b} The maps $\frac{B_{40}^{\prime}(\alpha,\beta)}{M_{4}}$ and
$\frac{B_{60}^{\prime}(\alpha,\beta)}{M_{6}}$ for Pr$^{3+}$:KY$_{3}$F$_{10}$ [19], $C_{4v}$ point symmetry; the
initial CFPs (in $cm^{-1}$) are $B_{40}=-1711$, $B_{60}=657$; $M_{4}=1862$, $M_{6}=713$; $x_{4}=-0.303$,
$x_{6}=-0.298$; the particular points are marked by the letters.

\noindent {\bf FIGS 6a, 6b} The maps $\frac{B_{40}^{\prime}(\alpha,\beta)}{M_{4}}$ and
$\frac{B_{60}^{\prime}(\alpha,\beta)}{M_{6}}$ for Pr$^{3+}$:Cs$_{2}$NaRCl$_{6}$ [20], $O_{h}$ point symmetry;
the initial CFPs (in $cm^{-1}$) are $B_{40}=2326$, $B_{60}=247$; $M_{4}=3045$, $M_{6}=699$; $x_{4}=0.5976$,
$x_{6}=-1.8708$; the particular points are marked by the letters.

\noindent {\bf FIGS 7a, 7b} The maps $\frac{B_{40}^{\prime}(\alpha,\beta)}{M_{4}}$ and
$\frac{B_{60}^{\prime}(\alpha,\beta)}{M_{6}}$ for Pr$^{3+}$ in Pr$_{2}$CuO$_{4}$ [21], $C_{4v}$ point symmetry;
the initial CFPs (in $cm^{-1}$) are $B_{40}=-2302$, $B_{60}=147$; $M_{4}=2991$, $M_{6}=1476$; $x_{4}=-0.830$,
$x_{6}=9.99$; the particular points are marked by the letters.

\noindent {\bf FIG.8} The ranges of the distinguished directions characterized by the maximal values of
$|B_{k0}^{\prime}|$ CFPs depending on the $|x|$ ratios for tetragonal ${\cal H}_{\rm CF}$'s ($|x_{4}|=0.5976$
and $|x_{6}|=1.8708$ are the cubic ratios, the $\beta$ angle between the distinguished direction and the
$z$-axis is given within the round brackets).


\newpage

\begin{center}
\includegraphics[width=21cm]{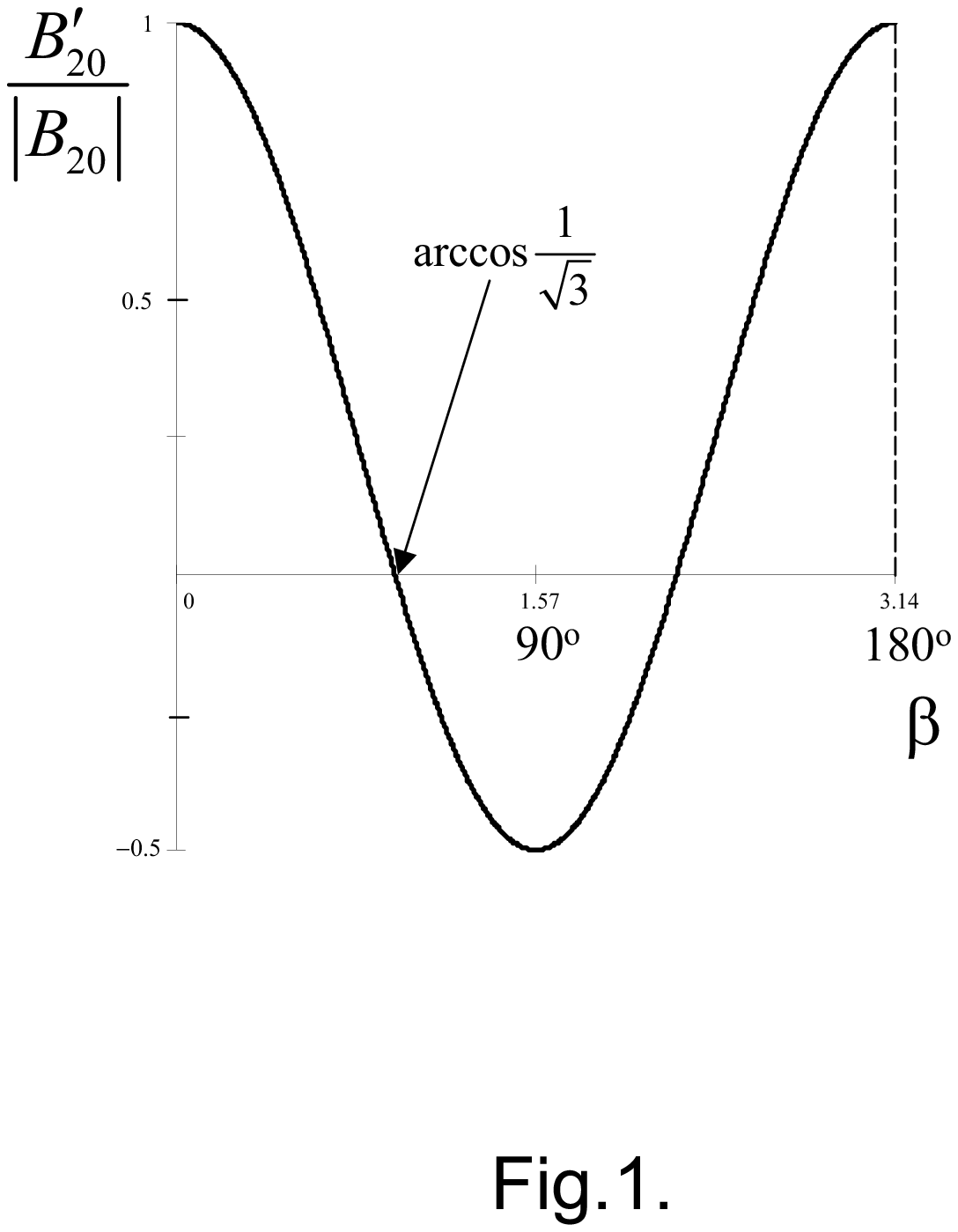}
\newpage
\includegraphics[width=19cm]{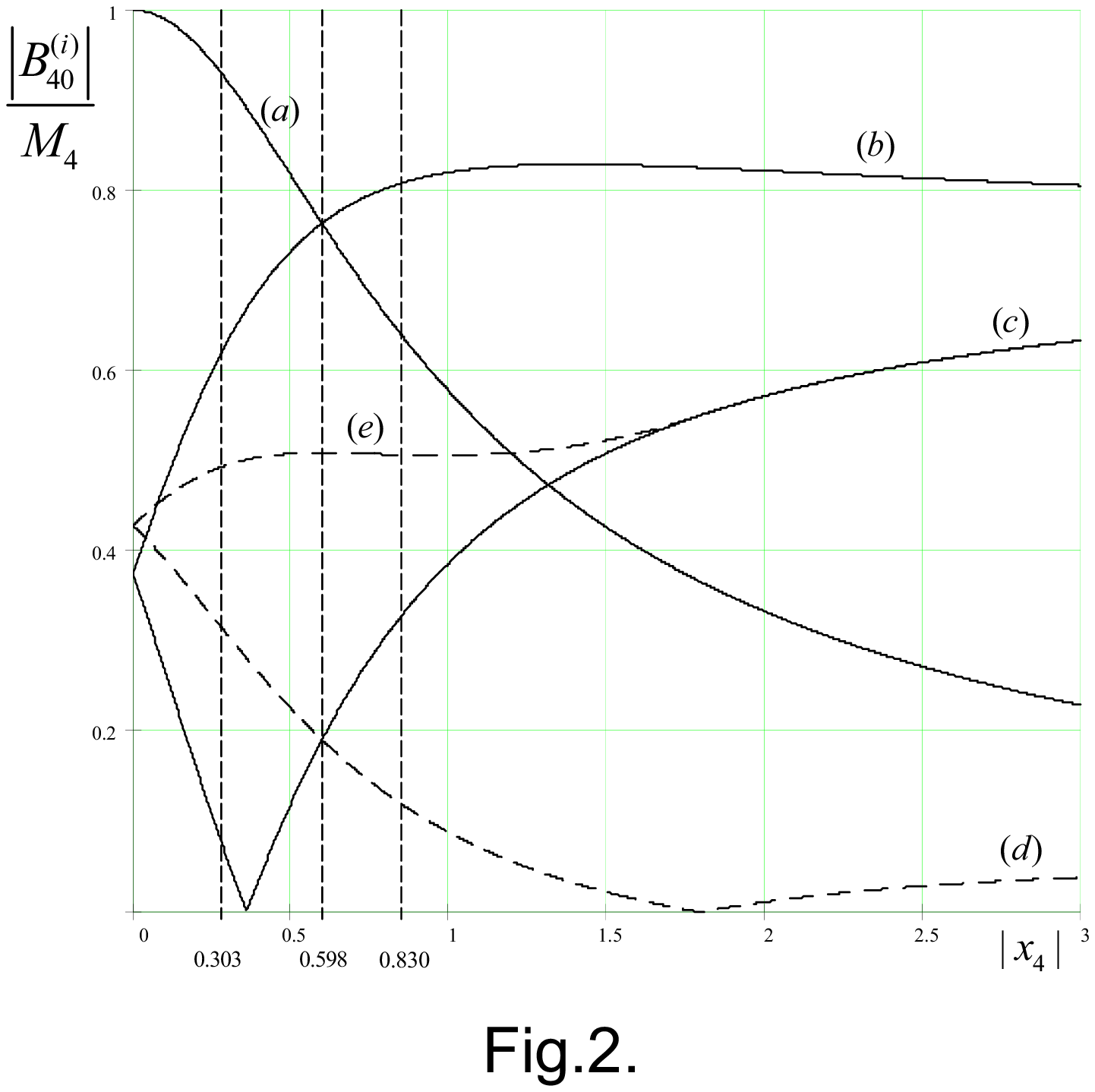}
\newpage
\includegraphics[width=19cm]{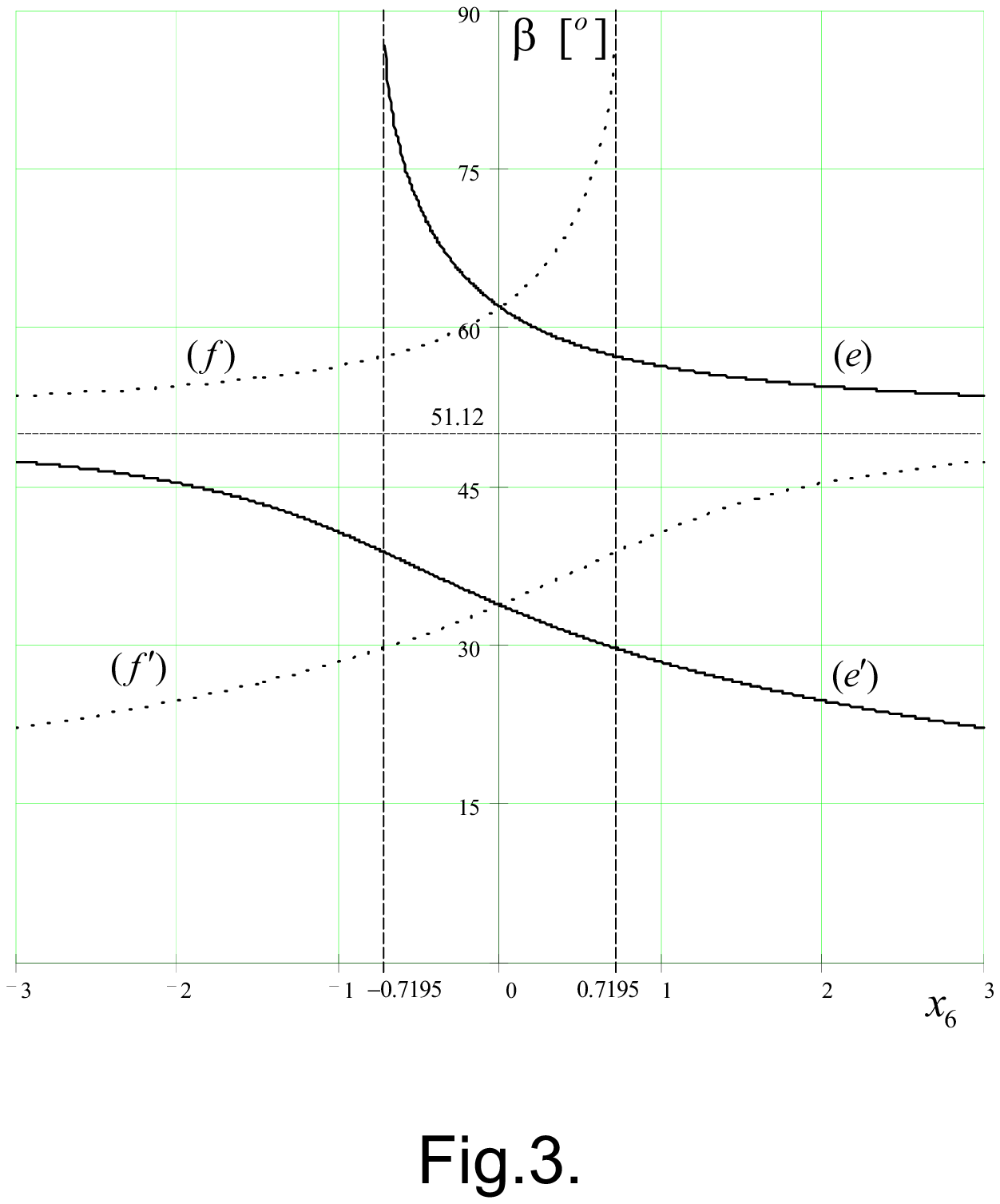}
\newpage
\includegraphics[width=19cm]{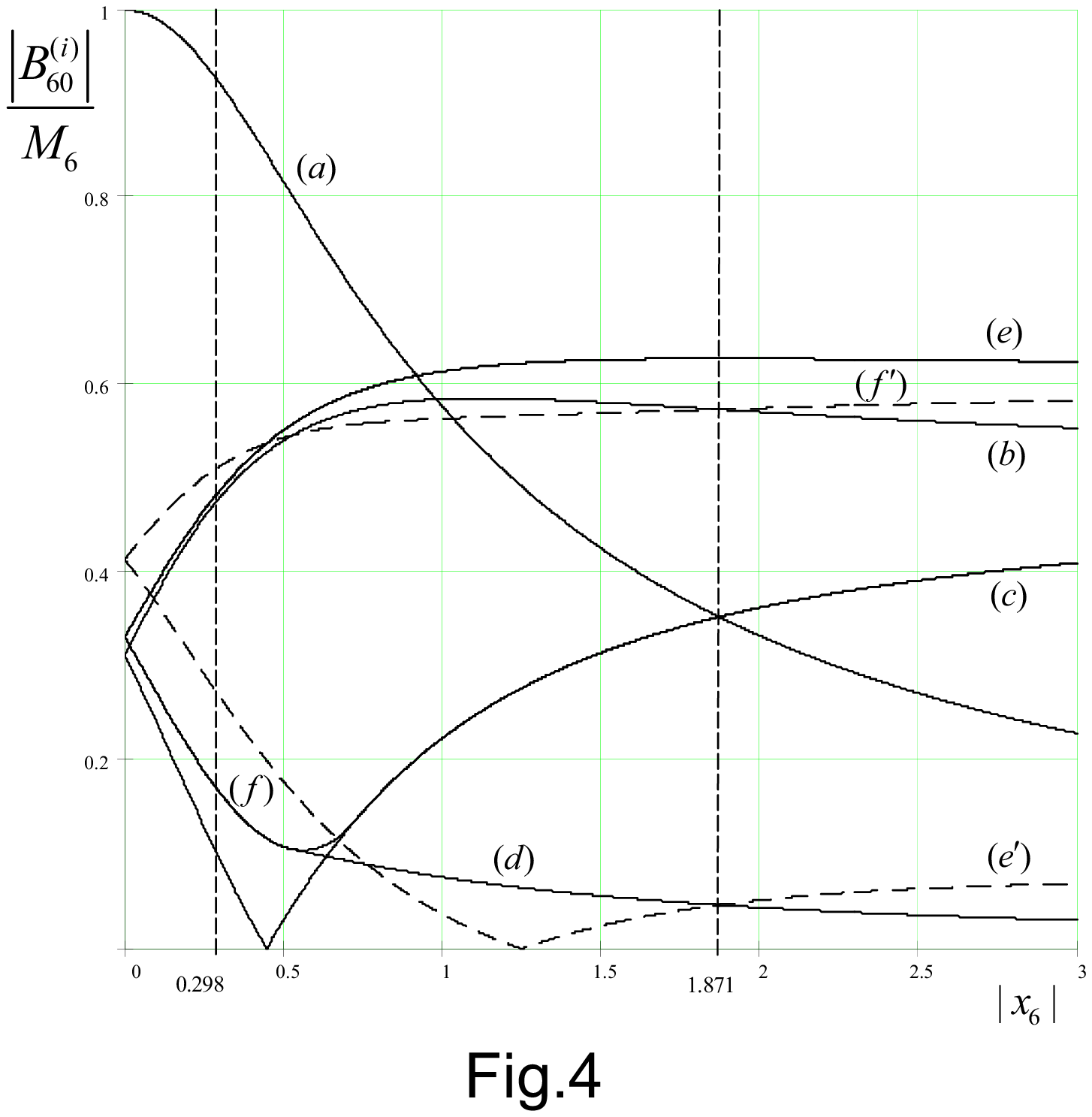}
\newpage
\includegraphics[width=20cm]{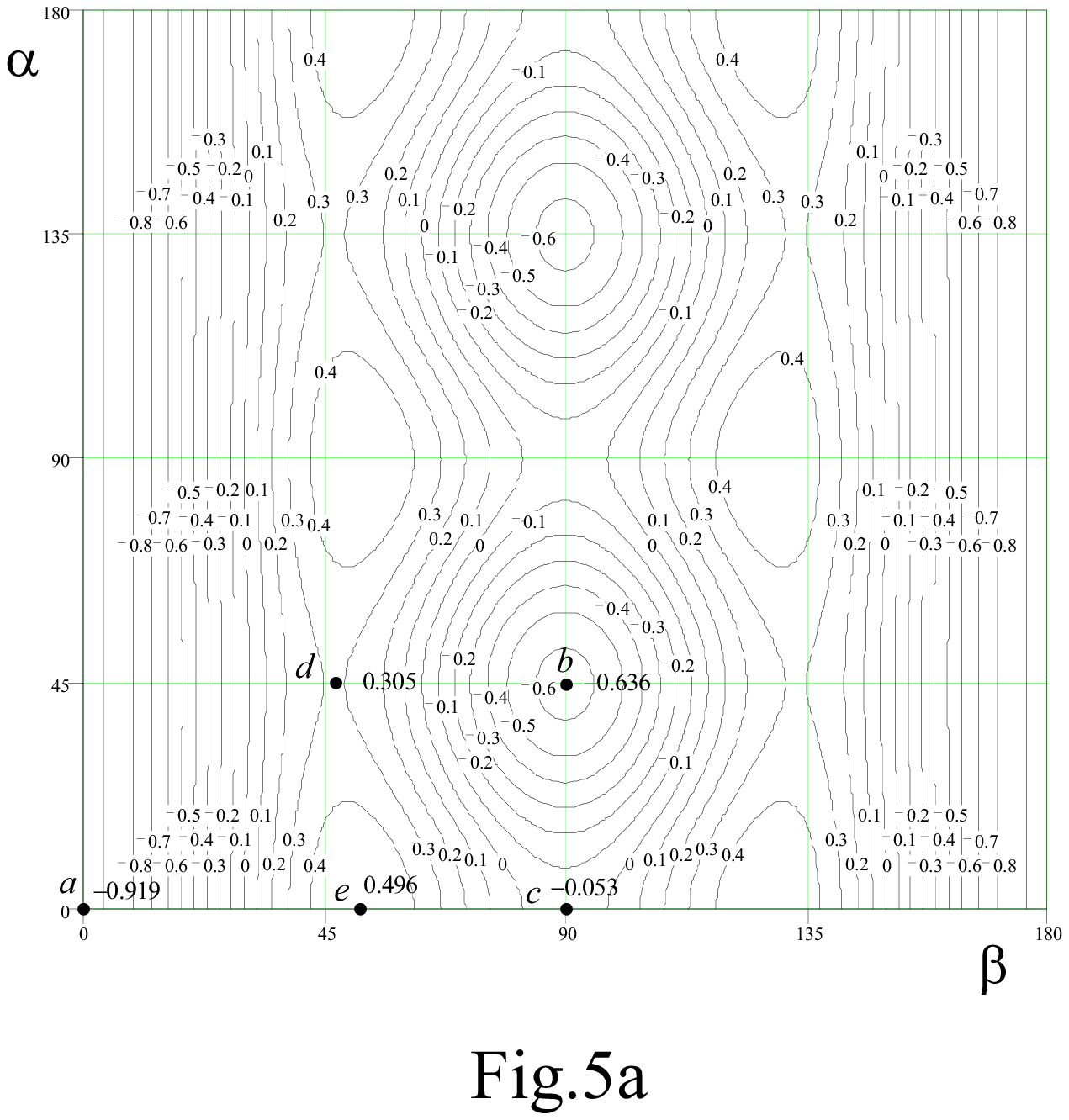}
\newpage
\includegraphics[width=19.5cm]{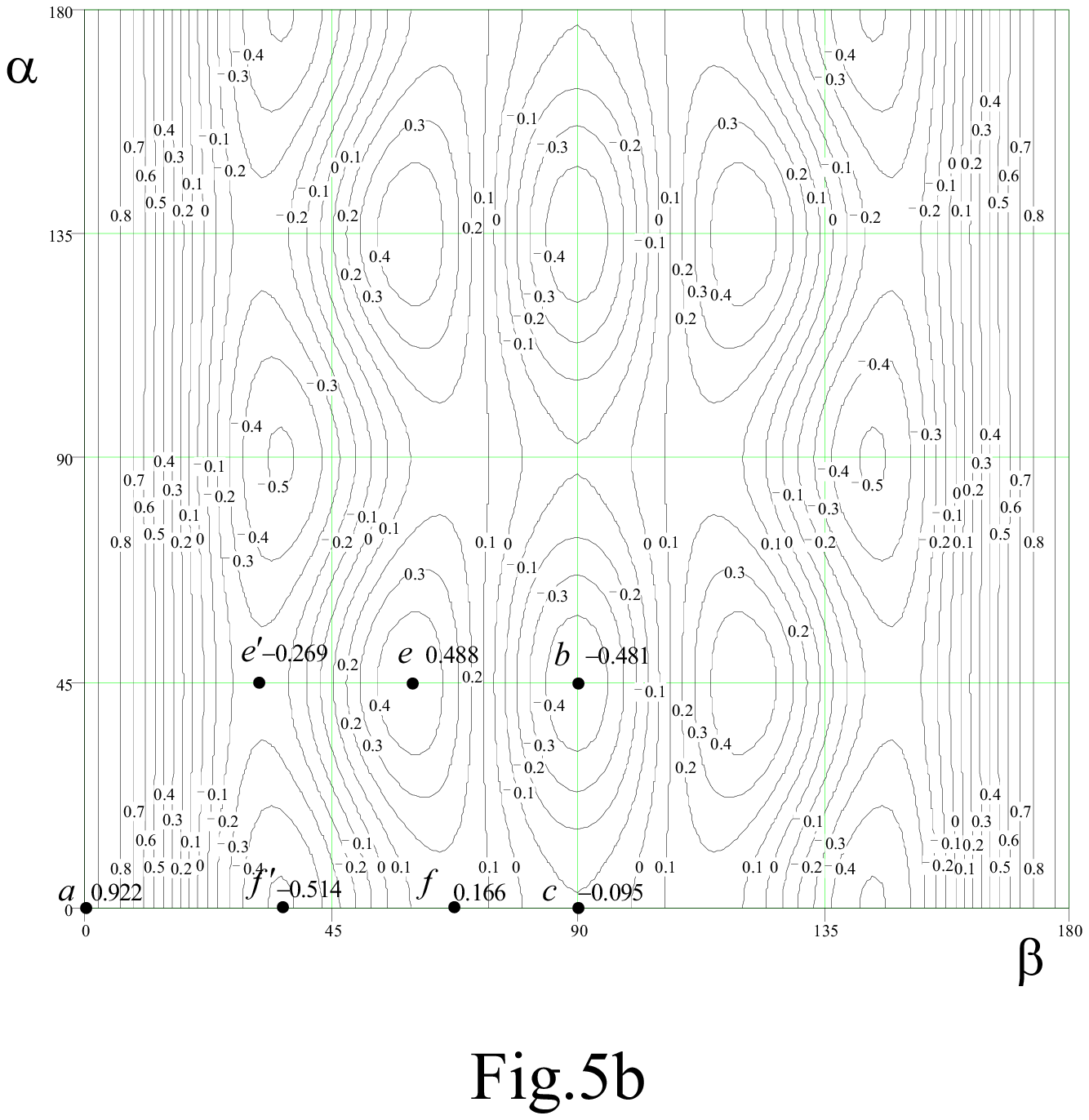}
\newpage
\includegraphics[width=18.5cm]{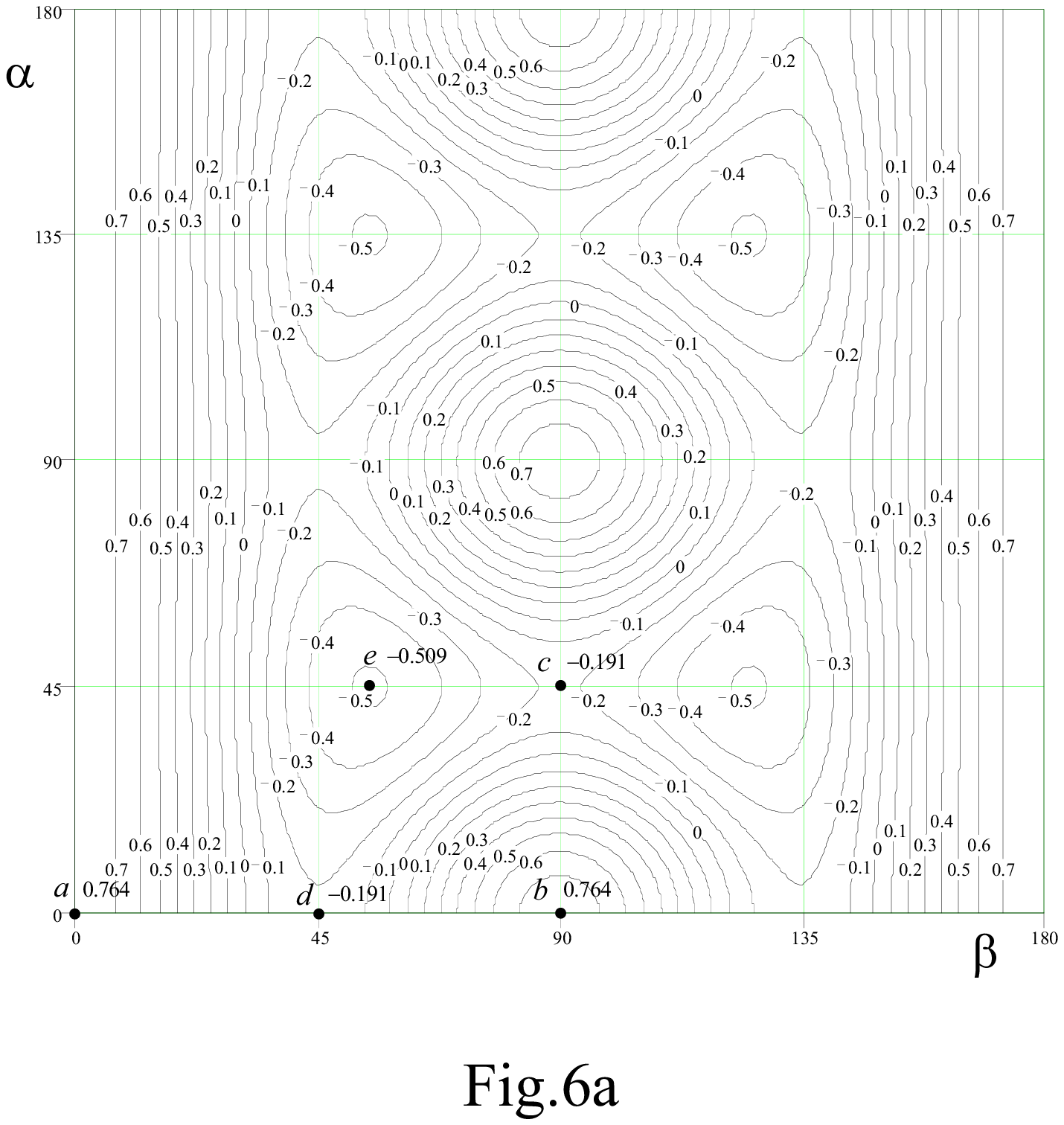}
\newpage
\includegraphics[width=19cm]{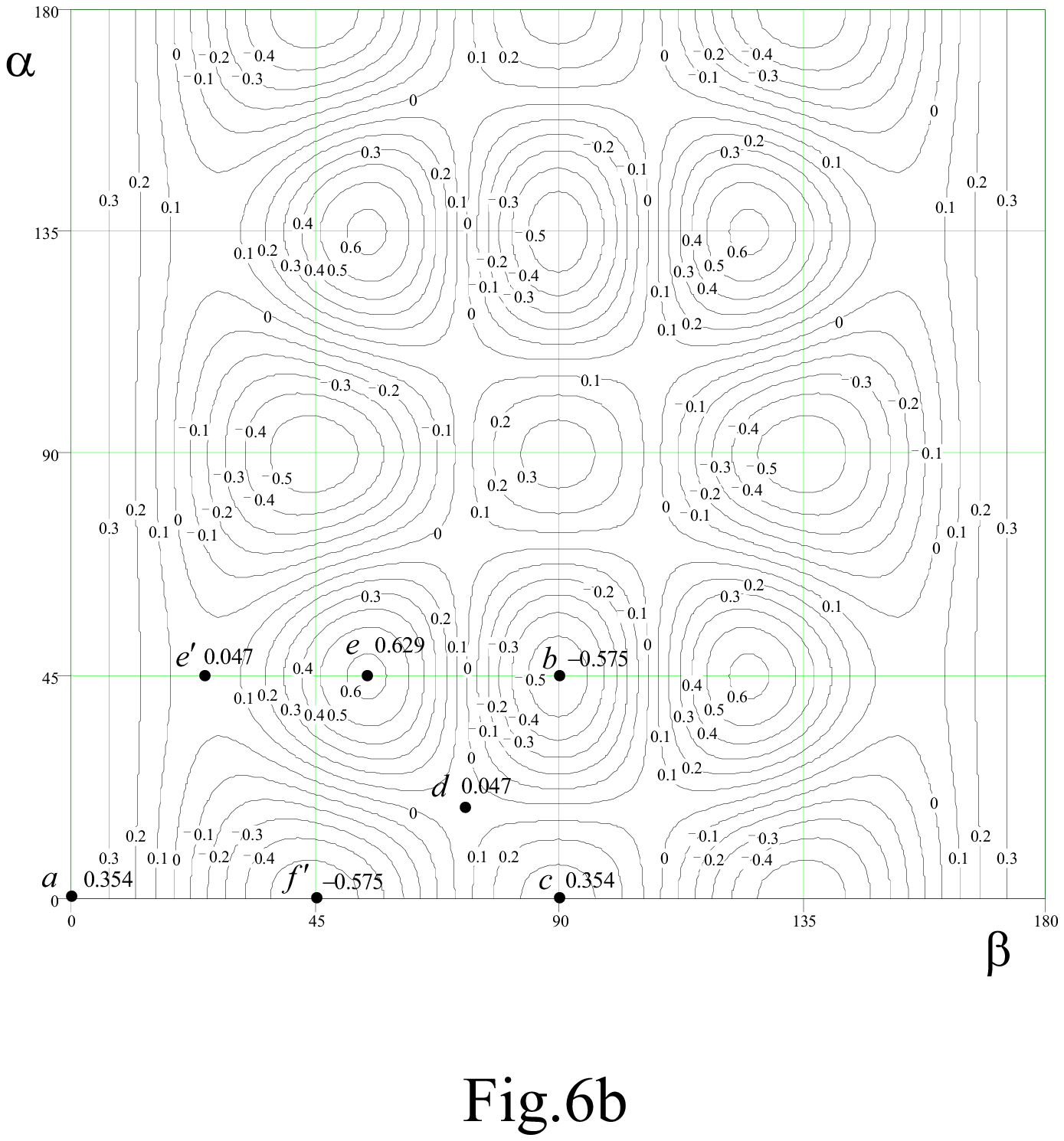}
\newpage
\includegraphics[width=19cm]{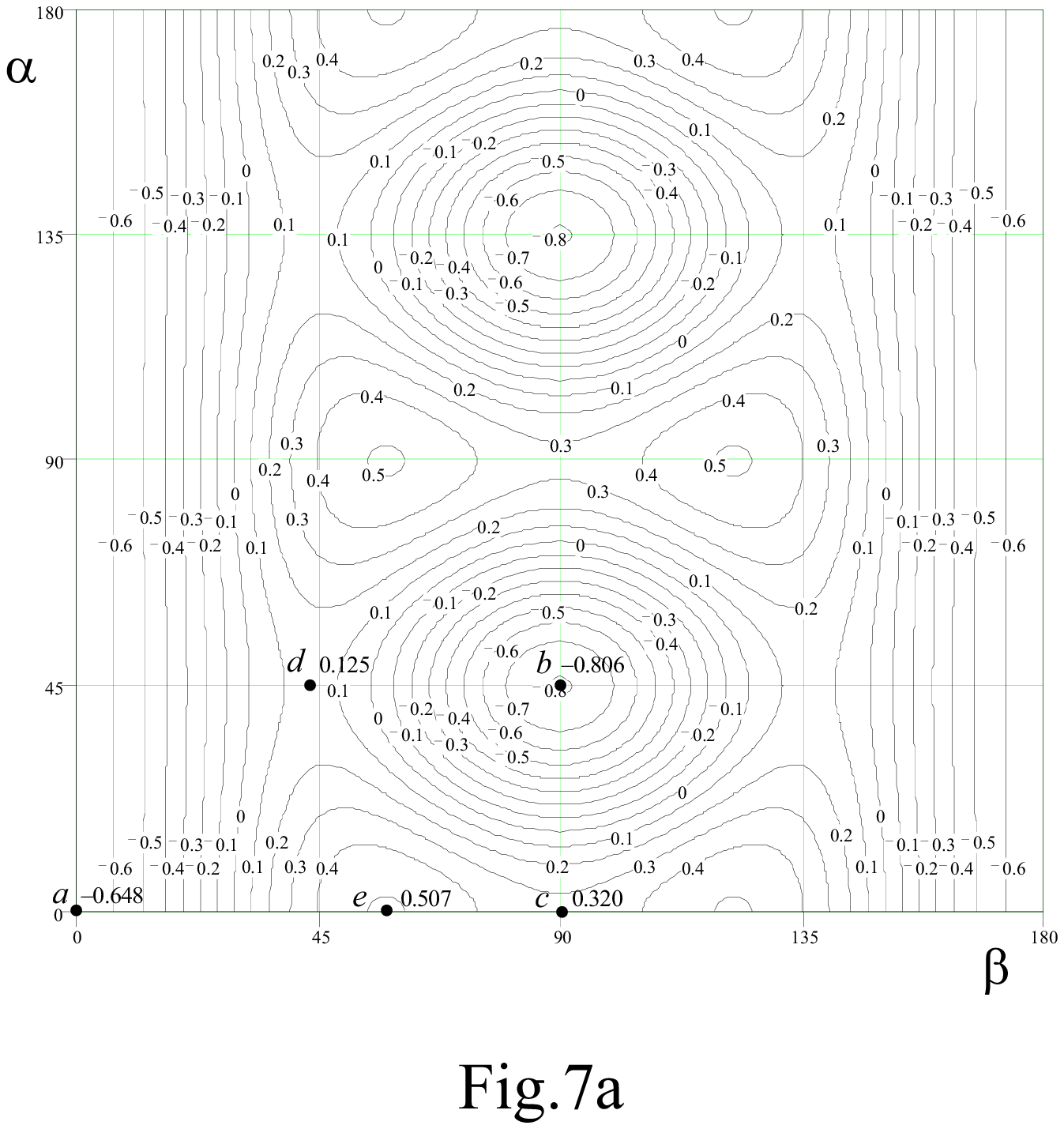}
\newpage
\includegraphics[width=18.5cm]{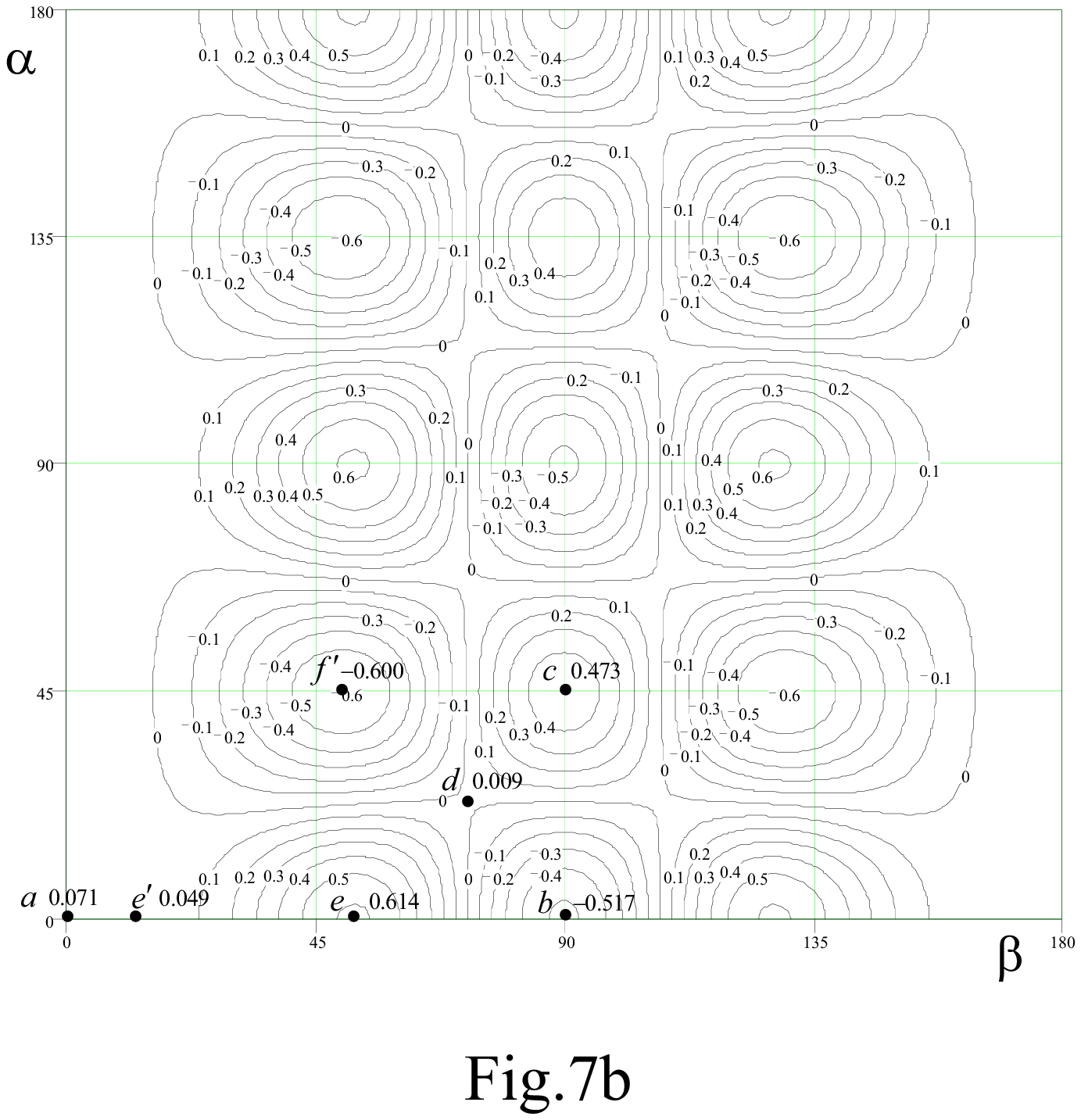}
\newpage
\includegraphics[width=18cm]{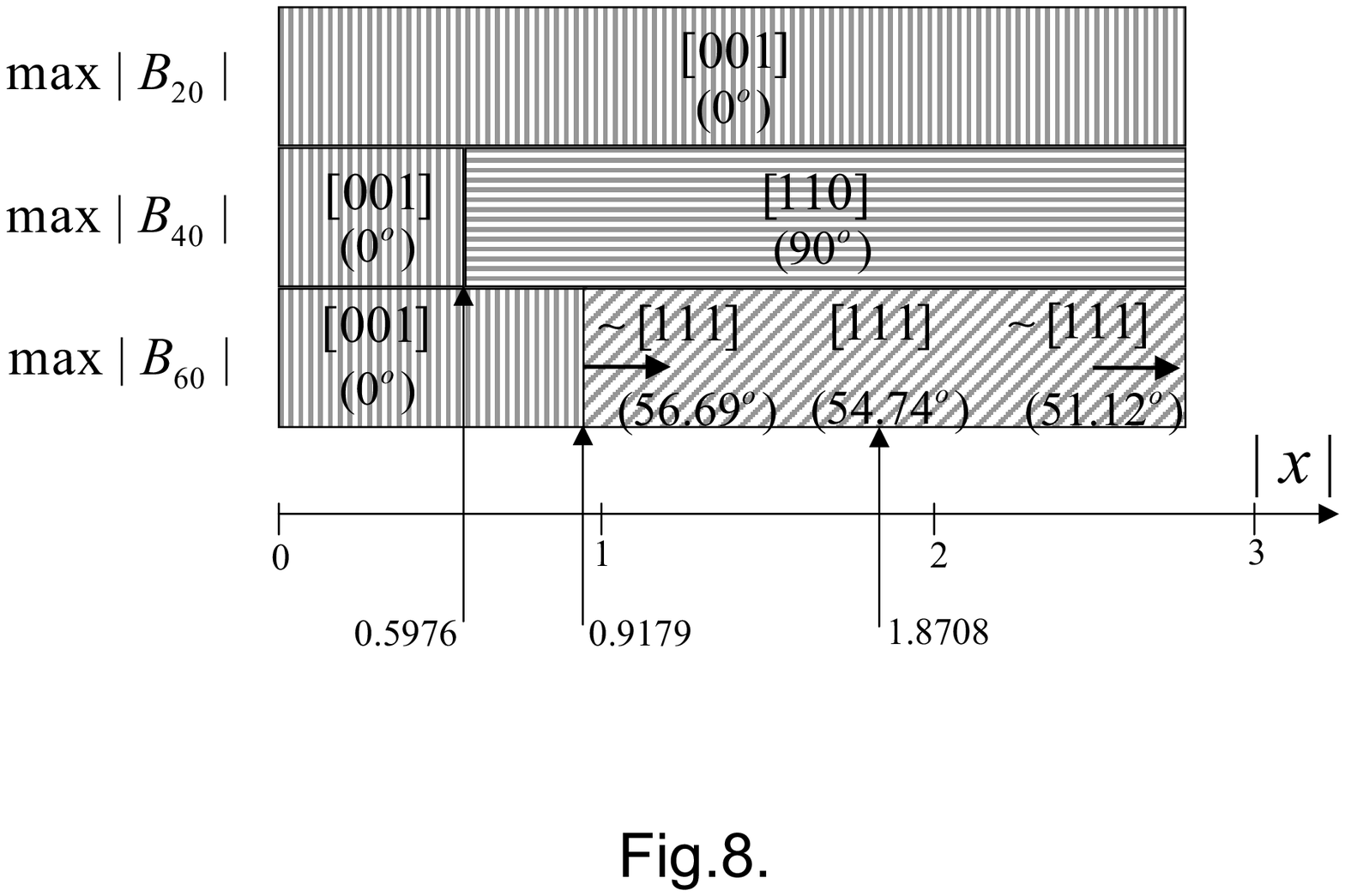}
\end{center}

\end{document}